\documentclass[aps,prd,twocolumn,floatfix,showpacs,superscriptaddress,nofootinbib]{revtex4-2}
\usepackage{amsmath,amsfonts,amssymb,bm}
\usepackage{graphicx}
\usepackage{color}
\usepackage{subfigure}
\usepackage{multirow}
\usepackage{textcomp}
\usepackage{slashed}
\usepackage{bbm}
\usepackage[bottom]{footmisc}
\definecolor{purple}{rgb}{0.5,0,0.5}
\definecolor{blue}{rgb}{0.0,0,1.0}
\usepackage[colorlinks=true, pdfstartview=FitV, linkcolor=purple, citecolor= purple, urlcolor=blue]{hyperref}

\newcommand{\al}{\alpha}

\newcommand{\beq}{\begin{equation}}
\newcommand{\eeq}{\end{equation}}
\newcommand{\ba}{\begin{array}}
\newcommand{\ea}{\end{array}}
\newcommand{\bea}{\begin{align}}
\newcommand{\eea}{\end{align}}
\newcommand{\bi}{\begin{itemize}}
\newcommand{\ei}{\end{itemize}}
\newcommand{\ben}{\begin{enumerate}}
\newcommand{\een}{\end{enumerate}}
\newcommand{\bc}{\begin{center}}
\newcommand{\ec}{\end{center}}
\newcommand{\bl}{\begin{flushleft}}
\newcommand{\el}{\end{flushleft}}
\newcommand{\br}{\begin{flushright}}
\newcommand{\er}{\end{flushright}}

\newcommand{\nn}{\nonumber \\}
\newcommand\Eqn[1]{Eq.~(\ref{#1})}  
\newcommand\Fig[1]{Fig.~\ref{#1}} 

\newcommand{\ie}{{i.e.}}

\newcommand\comment[1]{ \hbox{[{\it Comment suppressed here.}\/]} }
\newcommand\hide[1]{}
\newcommand{\skipover}[1]{}

\begin{document}
\title{Gravitational form factors of pseudoscalar mesons in a contact interaction}

\author{M. Atif Sultan}%
\email{atifsultan.chep@pu.edu.pk}
\affiliation{School of Physics, Nankai University, Tianjin 300071, China}
\affiliation{  Centre  For  High  Energy  Physics,  University  of  the  Punjab,  Lahore  (54590),  Pakistan}
\author{Zanbin Xing}%
\email{xingzb@mail.nankai.edu.cn}
\affiliation{School of Physics, Nankai University, Tianjin 300071, China}
\author{Kh\'epani Raya}%
\email{khepani.raya@dci.uhu.es}
\affiliation{Department of Integrated Sciences and Center for Advanced Studies in Physics, Mathematics and Computation, University of Huelva, E-21071 Huelva, Spain}
\author{Adnan Bashir}%
\email{adnan.bashir@umich.mx}
\affiliation{Department of Integrated Sciences and Center for Advanced Studies in Physics, Mathematics and Computation, University of Huelva, E-21071 Huelva, Spain}
\affiliation{Instituto de F\'isica y Matem\'aticas, Universidad Michoacana de San Nicol\'as de Hidalgo, Morelia, Michoac\'an 58040, M\'exico}
\author{Lei Chang}
\email{leichang@nankai.edu.cn}
\affiliation{School of Physics, Nankai University, Tianjin 300071, China}

\date{\today}
\begin{abstract}
Given the unique role played by the gravitational form factors (GFFs) in unraveling the internal mechanics of hadrons, we examine the GFFs of ground state pseudoscalar mesons $\pi$, $\eta_c$, $\eta_b$ and the hypothetical {\em strangeonium} $\eta_s(s\bar{s})$. We adopt the coupled framework of Dyson-Schwinger and Bethe-Salpeter equations within a contact interaction, and employ a novel approach to the dressed amputated meson-meson scattering amplitude which makes connection with the energy-momentum tensor and with the GFFs. The resulting GFFs fulfill the anticipated symmetry constraints. The corresponding charge and mass radii and the $D-$term are also computed. We show that the $D-$term for the pseudoscalar mesons is bounded within the $(-1, -1/3)$ range; these bounds correspond to the massless (chiral limit) and infinitely massive cases, respectively. Considering the current interest in the GFFs, understanding the \textit{D}-term of pseudoscalar mesons and their GFFs can provide an important first step for future endeavors in the field.
\end{abstract}

\maketitle
\section{introduction}
A meson composed of a valence quark and an antiquark of the same flavor is commonly denominated as a quarkonium. The first quarkonia states, the $\psi$ meson resonances, were observed in 1974, soon after the advent of quantum chromodynamics (QCD). 
Nowadays, experimental endeavors at LHC, Belle, \textit{BaBar}, and BES-III offer multiple pathways for the study of quarkonia systems, allowing to delve into various facets of QCD and hadronic physics (see, for example,\,\cite{Liu:2015zqa,Yuan:2019zfo,Chapon:2020heu}). Whereas for light pseudoscalar mesons like the pions, it is the dynamical breaking of chiral symmetry and its emergence as Nambu-Goldstone (NG) bosons which play a dictating role in understanding their internal dynamics\,\cite{Horn:2016rip}, for the heavier $\eta_c$ and $\eta_b$, the Higgs mechanism of mass generation is more dominant. The hypothetical {\em strangeonium}  meson
 $\eta_s(s\bar{s})$ separates the two regions of outright dominion. 
Charting out these limiting cases allows us to explore the transition between the two competing mass generating mechanisms\,\cite{Raya:2024ejx,Ding:2022ows}. Moreover, we get to probe contrasting facets of QCD at different energy scales because of the noticeably different spatial extent of the mesons under study and the distance dependence of the effective coupling strength between quarks. 
\par
Studying the partonic structure of hadrons has become increasingly popular for experimental and theoretical endeavors in the last decades. Some aspects, including the distribution of charge and magnetization are exposed via electromagnetic form factors (EFFs), whereas the gravitational form factors (GFFs) represent a complementary perspective by enabling access to the mechanical structure and mass distribution within  hadrons\,~\cite{Polyakov:2018zvc,Lorce:2018egm,Burkert:2023wzr}. Mathematically, the energy-momentum tensor (EMT) encodes the mass and spin decomposition of a given hadron, such that the definition of the GFFs emerges from matrix elements of the EMT~\cite{Pagels:1966zza,Novikov:1980fa}. Notably, GFFs also encompass an essential physical quantity known as the \textit{D}-term, which is associated with the mechanical stability of the hadron\,\cite{Polyakov:1999gs,Polyakov:2002yz,Perevalova:2016dln}. As will be discussed later in detail, the \textit{D}-term of NG bosons in the chiral limit is completely fixed by QCD symmetries\,\cite{Novikov:1980fa,Polyakov:1999gs,Polyakov:1998ze}. Beyond this limit, among various other theoretical frameworks, chiral perturbation theory ($\chi$PT) can estimate corrections to this value, but the domain of applicability of this approach completely leaves out quarkonia\,\cite{Donoghue:1991qv}. As discussed in Refs.~\cite{Novikov:1980fa,Voloshin:1980zf,Xu:2024hfx}, GFFs can provide a proper depiction of these systems and describe their hadronic decays. 

From an experimental point of view, a direct access to the GFFs would require the interaction of the hadron with a gravitational probe\,\cite{Kobzarev:1962wt,Voloshin:1982eb}. Therefore, naturally enough, one has to resort to indirect means to extract these quantities, for example, through electromagnetic processes which also provide insight into how these quantities are related to other quantities such as generalized distribution amplitudes and generalized parton distributions (see \emph{e.g.}\,\cite{Mezrag:2022pqk}); within this perspective, some explorations on the pion and proton GFFs have already been carried out in different experiments (see \textit{e.g.} Refs.~\cite{Kumano:2017lhr,Burkert:2018bqq,Burkert:2021ith,Burkert:2018bqq,Dutrieux:2021nlz,Burkert:2023atx,Duran:2022xag}). The precision of such efforts is expected to improve in the forthcoming era of electron-ion colliders~\cite{Accardi:2012qut,Anderle:2021wcy}. Moreover, lattice QCD (lQCD) data for the pion and the proton GFFs have recently become available near the physical pion mass, \emph{i.e.} $m_\pi^{lat}=170$ MeV\,\cite{Hackett:2023nkr,Hackett:2023rif}. The availability of improved lQCD data opens new avenues for theoretical investigations and subsequent experimental validations. Alternatively, there are theoretical approaches which are more amicable towards working in the chiral limit than lQCD and can provide robust predictions and  explanations which can simultaneously describe the internal structure of mesons, baryons, and other bound states resulting from QCD interactions\,\cite{Wang:2024sqg}. These studies can potentially deepen our understanding of QCD\,\cite{Ding:2022ows,Raya:2024ejx}. With that in mind, we expect that the upcoming precise experimental results and lQCD data for quarkonia GFFs will be able to verify these theoretical predictions, particularly regarding the value of the \textit{D}-term.
In this manuscript, we take one particular alternative route based upon the continuum description of QCD and hadron physics, namely the field theoretic fundamental equations 
of motions dubbed as Dyson-Schwinger Equations (DSEs).

Due to its intrinsic capacity to capture both the perturbative and non-perturbative aspects of QCD within one single formalism, while simultaneously preserving the key symmetries of the  theory, the DSEs approach provides a reliable and  robust framework to study the internal structure of mesons and baryons,\,\cite{Qin:2020rad,Eichmann:2016yit}. Earliest studies of heavy quarkonia through DSEs can be found in Refs.~\cite{Jain:1993qh,Krassnigg:2004if,Bhagwat:2004hn,Maris:2005tt,Bhagwat:2006xi,Krassnigg:2009zh,Souchlas:2010zz,Blank:2011ha,Rojas:2014aka,Fischer:2014cfa,Ding:2015rkn,Raya:2016yuj,Raya:2017ggu,Xu:2019ilh}. More than a decade ago, an alternative approach to the numerically involved DSE calculations was developed. That is the symmetry preserving  contact interaction (CI) introduced in~\cite{Gutierrez-Guerrero:2010waf,Roberts:2010rn}. It captures the infrared properties of QCD rather well. This model was initially employed to study the properties of the pion in an amicable algebraic framework, capable of preserving the important infrared features of QCD, {\em i.e.}, dynamical chiral symmetry breaking and confinement. Its wide range of applications cover the study of static and dynamic properties of hadrons in vacuum, such as mass spectrum\,\cite{Roberts:2011cf,Chen:2012qr,Bedolla:2015mpa,Yin:2019bxe,Gutierrez-Guerrero:2021rsx}; electromagnetic, electroweak and GFFs\,\cite{Gutierrez-Guerrero:2010waf,Roberts:2010rn,Roberts:2011wy,Wilson:2011aa,Segovia:2014aza,Cheng:2022jxe,Xing:2022mvk,Xing:2022jtt,Hernandez-Pinto:2023yin,Zamora:2023fgl,Dang:2023ysl,Raya:2021pyr};
and even generalized parton distributions~\cite{Zhang:2020ecj,Xing:2023eed}; on the other hand, it has also proven useful for investigating in-medium properties such as the phase diagram of QCD~\cite{Ahmad:2020jzn,Wang:2013wk,Serna:2016ifh}. Of course, the list of works listed herein is only representative and does not include the entire assortment of contributions.  Nevertheless, it is worth mentioning that the CI is easy to implement and the explorations within it provide valuable benchmarks for more sophisticated treatments of QCD and experiment; to a large extent, because the strengths and limitations of the model are clearly discernible. Thus, it has developed as a powerful tool to investigate the hadronic properties.  Considering the discussion so far as a satisfactory justification, we extend the CI analysis in Ref.\,\cite{Xing:2022mvk} on the GFFs of the pion to include $\eta_c$ and $\eta_b$, and the hypothetical {\em strangeonium} $\eta_s(s\bar{s})$.

The manuscript is organized as follows: in Sec.\,\ref{sec::gff} general features of the GFFs and its calculation within the present CI model, based upon the DSE approach, are presented. The corresponding numerical results are provided in detail and are discussed in Sec.\,\ref{sec::results}. Finally, summary and conclusion are presented in Sec.\,\ref{sec::summary}, along with  perspectives and possibilities for future work. 

\section{Gravitational form factors}\label{sec::gff}
\subsection{General aspects}
The expectation value of the EMT for a spin-0 meson is expanded out as follows in terms of covariant tensors:\footnote{We employ a Euclidean metric with $\{\gamma_\mu,\gamma_\nu\} = 2\delta_{\mu\nu}$; $\gamma_\mu^\dagger = \gamma_\mu$; $\gamma_5= \gamma_4\gamma_1\gamma_2\gamma_3$; and $a \cdot b = \sum_{i}^{4} a_i b_i$. The isospin symmetric limit $m_{u/d}=m_l$ is considered throughout this work.}
\begin{eqnarray}
\label{eq:defMGFF}
&& \mathcal{M}^{H}_{\mu\nu}(Q^2) \hspace{-0.5mm} = \hspace{-0.5mm} 2P_\mu P_\nu A^{H}(Q^2) \hspace{-0.5mm} + \hspace{-0.5mm} \frac{1}{2}\left(Q_\mu Q_\nu-Q^2\delta_{\mu\nu}\right)
\hspace{-1mm} D^{H}(Q^2) \nonumber \\
&& \hspace{16mm} +2m_H^2\delta_{\mu\nu}\bar{c}^{H}(Q^2)\,,
\end{eqnarray}
where $Q$ is the external momentum of the probe and $P$ is the average between the incoming and outgoing momenta, such that $P^2=-m_H^2-Q^2/4$, with $m_H$ being the meson mass; the script $H=\{PS, S\}$ denotes the pseudoscalar ($PS$) and scalar ($S$) mesons. The Lorentz scalar functions $\mathcal{G}^H(Q^2)=\{A^{H}(Q^2),D^{H}(Q^2),\bar{c}^{H}(Q^2)\}$ correspond to the GFFs.

It is crucial to recognize that the GFFs are  constrained by symmetry principles. Firstly, EMT conservation implies $Q_\mu \mathcal{M}^H_{\mu\nu}(Q^2)= Q_\nu \mathcal{M}^H_{\mu\nu}(Q^2) = 0$, which results in $\bar{c}^H(Q^2)=0$. On the other hand, for the  zero-momentum transfer, one expects~\cite{Kobzarev:1962wt,Pagels:1966zza}:
\begin{eqnarray}
\lim_{Q^2\rightarrow0} A^H(Q^2)=1\,,\quad \lim_{Q^2\rightarrow0} Q^2 D^H(Q^2)=0\,.
\end{eqnarray}
The equation on the left stems from the mass normalization, whereas the second one encompasses the equilibrium of forces within the bound-state. It is known as the von Laue condition\,\cite{Polyakov:2018zvc,Hudson:2017xug}. The $D-$term is also defined in this domain, via:
\begin{eqnarray}
    \lim_{Q^2\rightarrow0} D^H(Q^2):=D^H\,.
\end{eqnarray}
There are no established constraints on $D^H$ for arbitrary hadrons, with the exception of the one arising from the soft pion theorem. It entails the following result for the $PS$ mesons in the chiral limit~\cite{Novikov:1980fa,Polyakov:1999gs,Polyakov:1998ze}:
\begin{equation}
\label{eq:sptc}
    D^{PS}\overset{m_{\text{PS}}^2=0}{=} D^{PS}_{c.l.}:= -1\;.
\end{equation}
The nature of this condition is tightly interwoven with dynamical chiral symmetry breaking\,\cite{Mezrag:2014jka,Xu:2023izo}. The mass dependence of the first correction to it can be formally derived from $\chi$PT for the pseudo Nambu-Goldstone (NG) bosons\,($\pi,\,K,\,\eta$ mesons),\,\cite{Donoghue:1991qv,Polyakov:2018zvc,Hudson:2017xug}. The tendency in such a case is to reduce its absolute value proportionally to the quadratic NG boson mass: $D^{NG}=-1+\mathcal{O}(m_{NG}^2)$. 
\par
Before proceeding to detail the computational framework, let us recall that the total GFFs of a given hadron $H$ result from adding up the individual contributions of each parton type-$a$, namely $\mathcal{G}^H(Q^2)=\sum_{a}\mathcal{G}^{H}_{a}(Q^2)$. In the present approach, the entirety is contained within the fully dressed valence quarks such that in the $f$ flavor symmetric systems:
\begin{equation}
    \mathcal{G}^H(Q^2) = 2 \mathcal{G}^H_f(Q^2)\,.
\end{equation}
 Thus, in what follows, we shall ignore the quark flavor label and only specify the meson type when necessary.

\subsection{The scattering amplitude in the CI model}
When studying meson EFFs within the DSE formalism, it is common to appeal to the impulse approximation, which results in the computation of a triangle diagram that incorporates a fully-dressed quark-photon vertex (QPV), see \emph{e.g.} Refs.\,\cite{Gutierrez-Guerrero:2010waf,Roberts:2010rn,Roberts:2011cf}.  An analogous approach for the GFFs requires the QPV to be replaced by the quark-tensor vertex (QTV) whose rigorous construction is naturally challenging\,\cite{Xu:2023izo}. Here we proceed differently, taking into account the novel perspective presented in Ref.~\cite{Xing:2022mvk} for the calculation of the pion GFFs. This features a straightforward derivation of the amputated $HH$ scattering amplitude for the EMT. The reasons for this choice, and its implementation, are discussed below.

In the present hadronic scale picture, where the hadron properties are entirely contained within its fully-dressed valence quarks, it is sufficient to consider the EMT valence-quark contribution to Eq.\,\eqref{eq:defMGFF}, which reads:\begin{align}\label{eqn::emt}
\mathcal{M}_{\mu\nu}(Q^2)=&2N_c\text{tr}\int_q \Tilde{\gamma}^{G}_{\mu\nu}(q+k,q+p) S(q+p)\nn
&\times i^2F(q,p,-k) S(q+k)\,.
\end{align}
This expression involves not only the dressed quark propagator, $S(q)$, but also the QTV, $\Tilde{\gamma}^{G}_{\mu\nu}$, and the dressed $HH$ scattering amplitude, $F(k,P_1,P_2)$. The quark propagators is obtained through its DSE, commonly denominated gap equation; in the CI model:
\begin{equation}
\label{eq:quarkPropCI}
    S^{-1}(p)=i \gamma \cdot p+m+\frac{4}{3 m_G^2} \int_{q}  \gamma_\mu S(q) \gamma_\mu\;,
\end{equation}
which yields the following convenient expression for the quark propagator,
\begin{equation}
    \label{eq:quarkPropCIgen}
    S^{-1}(p)=i\gamma\cdot p + M\,.
\end{equation}
Herein, $m_G$ is an infrared mass scale, $m$ the current quark mass, and $M$ is the resulting momentum-independent dressed mass; $\int_q:=\frac{d^4q}{(2\pi)^4}$ stands for a Poincar\'{e} covariant integral.
Note that Eq.\,\eqref{eq:quarkPropCI} is divergent and requires the implementation of a regularization scheme; and, in general, the CI model leads to integrals of the type:
\begin{equation}\begin{aligned}
I_{-2\alpha}(s)& :=\int_{q}\frac{1}{(q^{2}+s)^{\alpha+2}}\,,  \\
I_{-2\alpha}^{\mu\nu}(s)&:=\int_q\frac{q_{\mu}q_{\nu}}{(q^2+s)^{\alpha+3}}\,.
\end{aligned}\end{equation}
Employing the symmetry-preserving regularization procedure introduced in Ref.~\cite{Xing:2022jtt}, these integrals are regularized as follows: 
\begin{equation}\begin{aligned}
I_{-2\alpha}(s)\to I_{-2\alpha R}(s)& =\int_{\tau_{uv}^{2}}^{\tau_{ir}^{2}}d\tau\frac{\tau^{\alpha-1}}{\Gamma(\alpha+2)}\frac{e^{-\tau s}}{16\pi^{2}},\\
I_{-2\alpha}^{\mu\nu}(\mathcal{M}^2)\to I_{-2\alpha R}^{\mu\nu}(s)&=\frac{\Gamma(\alpha+2)}{2\Gamma(\alpha+3)}\delta_{\mu\nu}I_{-2\alpha R}(s)\,,
\end{aligned}\end{equation}
where $\tau_{ir}^{-1}=\Lambda_{ir}$ whereas $\tau_{uv}^{-1}=\Lambda_{uv}$ and serve as infrared (IR) and ultraviolet (UV) cutoffs, respectively. A nonzero value of $\Lambda_{ir} \sim \Lambda_{\text{QCD}}$ supports a confinement-compatible picture by preventing quark production thresholds, while $\Lambda_{uv}$ is essential and has a dynamic role since the CI does not constitute a renormalizable theory.

It is worth pointing out that the direct interaction between the tensor probe and gluons is possible, and should be considered to ensure the conservation of EMT both at the quark and meson levels. The gluon-tensor interaction will generate new contributions to ensure $\bar{c}(Q^2)=0$. However, in the CI model, it neither affects $A(Q^2)$ nor $D(Q^2)$ GFFs~\cite{Xing:thesis}. 
Along with the meson EMT conservation, the gluon-tensor coupling is also crucial to ensure that the QTV obeys its own Ward-Green-Takahashi identity\,\cite{Brout:1966oea}. Furthermore, the contributions from such interaction will modify the in-homogeneous part of the QTV Bethe-Salpeter equation (BSE) and, in CI model, producing\,\cite{Xing:thesis}:
\begin{equation}\label{eqn::bareQTV}
\Tilde{\gamma}^{G}_{\mu\nu}(k^+,k^-)=i\gamma_\mu\frac{k^+_\nu+k^-_\nu}{2}-\delta_{\mu\nu}\frac{S^{-1}(k^+)+S^{-1}(k^-)}{2}\,,
\end{equation}
where $k^\pm=k+P/2$. Notably, this structure resembles that of its tree-level counterpart\,\cite{Xing:2022mvk}, but with the dressed quark propagators. This indicates that, although the gluon dynamics (and consequently the gluon-tensor interaction) is cloaked by the CI, it is reflected in the dynamically enhanced dressed quark mass. For that reason, in what follows, we will refer to $\Tilde{\gamma}^{G}_{\mu\nu}(k^{+},k^{-})$ as the \emph{effective bare QTV}. Note that the component proportional to $\delta_{\mu\nu}$ is crucial in guaranteeing WGTI is satisfied and it only impacts $\bar{c}(Q^2)$. 

The third element entering Eq.\,\eqref{eqn::emt},
the dressed amputated $HH$ amplitude $F(k,P_1,P_2)$, satisfies an inhomogeneous BSE. In the ladder approximation, this reads\,\cite{Xing:2022mvk,Xing:2023eed}:
\begin{eqnarray}\label{eqn::pipiamplitude}
&&F(k,P_1,P_2)=F_0(k,P_1,P_2)+\Sigma^F(k,P_1,P_2)\,,
\end{eqnarray}
with the bare $HH$ amplitude 
\begin{eqnarray}\label{eqn::bamp}
F_0(k,P_1,P_2)=\Gamma_{H}(P_1)S(k)\Gamma_{H}(P_2)\,,
\end{eqnarray}
and the self-energy term 
\begin{align}\label{eqn::sigma}
\Sigma^F(k,P_1,P_2)=&-\frac{4}{3m_G^2}\int_q \gamma_{\al}S(q+P_1)\nn
&\times F(q,P_1,P_2)S(q-P_2)\gamma_{\al}\,.
\end{align}
The CI model gives $\Sigma^F(k,P_1,P_2)$ a plainly simple tensor decomposition, completely determined by the structures $\mathbbm{1}$ and $\gamma \cdot K$\,\cite{Xing:2022mvk}; here $K = (P_1-P_2)/2$, with $P_{1,2}$ the meson total momenta. Focusing on the inhomogeneous term, \emph{i.e.},~\Eqn{eqn::bamp}, $\Gamma_{H}(P)$ denotes the meson Bethe Salpeter Amplitude (BSA). This is obtained from the corresponding BSE:
\begin{eqnarray}
\label{eq:CIinMRL}
\Gamma_H(P)=-\frac{4}{3m_{G}^{2}}\int_{q}\left[\gamma_{\mu}S_{g}(q)\Gamma_H(q;P)S_f(q-P)\gamma_{\mu} \right]\,.\hspace{0.6cm} 
\end{eqnarray}
Given the momentum independent nature of the CI, the BSAs acquire a very simple structure, independent of the relative momentum between the valence quark and antiquark. In particular, for pseudoscalar and scalar mesons, respectively:
\begin{eqnarray}
    \Gamma_{PS}(P)&=&\gamma_5\left[i E_{PS}(P)+ \frac{1}{\bar{M}}\gamma \cdot P F_{PS}(P)\right]\,,\\
    \Gamma_{S}(P)&=& \mathbbm{1}E_{PS}(P)\,,\label{eq:ScalarBSA}
\end{eqnarray}
with $\bar{M}={2M_g M_f}/{(M_g+M_f)}$, which is simply $\bar{M}=M$ in the case of identical flavors or equal masses. The choice of parameters and the generated masses will be presented in the next section.
\par
\begin{figure}
\includegraphics[width=8.6cm]{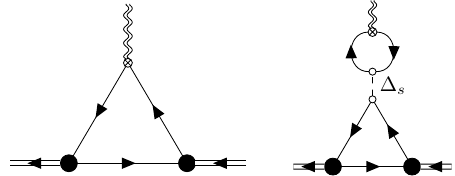}
\caption{\emph{Left:} $\mathcal{M}_{1\mu\nu}$, the impulse approximation piece with the effective bare QTV. \emph{Right:} $\mathcal{M}_{2\mu\nu}$, showing contribution from the  scalar meson pole. Solid, double-solid, double-wavy and dashed lines represent quark, pion, tensor probe, and the effective scalar propagator $\Delta_S$, respectively. Filled, crossed and empty circles represent PS/S meson BSA, effective bare QTV $\Tilde{\gamma}^{G}_{\mu\nu}$, and bare quark-scalar vertex, respectively.}
\label{fig::feyngff}
\end{figure}

  
Once the components entering Eq.\,\eqref{eqn::emt} have been determined, as detailed in Ref.~\cite{Xing:2022mvk}, it is convenient to cast $\mathcal{M}_{\mu\nu}^H(Q^2)$ as follows:
\begin{eqnarray}
\mathcal{M}^{H}_{\mu\nu}(Q^2)&=&\mathcal{M}^{H}_{1\mu\nu}(Q^2)+\mathcal{M}^{H}_{2\mu\nu}(Q^2)\,.
\end{eqnarray}
Here, $\mathcal{M}_{1\mu\nu}^H$ corresponds to the so-called impulse approximation involving the insertion of the effective bare QTV, $\Tilde{\gamma}^{G}_{\mu\nu}$, while $\mathcal{M}^{H}_{2\mu\nu}$ encodes the contributions arising from the scalar meson poles. Graphical representations of both elements are shown in \Fig{fig::feyngff}, and the explicit mathematical structures can be found in Ref.~\cite{Xing:2022mvk}. 

It should be emphasized that the contribution of $\mathcal{M}^{H}_{2\mu\nu}$ is essential to satisfying the soft-pion theorem\,\cite{Xing:2023eed,Xing:2022mvk}; namely, the corollary in Eq.\,\eqref{eq:sptc}. This is achieved through the manifestation of the pole corresponding to the scalar meson, whose effective propagator is found to be:
\begin{equation}
\label{eq:DefDelS}
\Delta_{S}(Q^2)=\frac{1}{3 m^2_G/4+f_S(Q^2)}\,,
\end{equation}
where $f_S$ is the scalar vacuum polarization at the one-loop level\,~\cite{Xing:2022mvk}. Given the simplicity and constancy of the scalar meson BSA within the CI,\,Eq.\,\eqref{eq:ScalarBSA}, it is anticipated that the $Q^2$ evolution of the corresponding FF would be harder than the exact result. This situation was ameliorated in Ref.~\cite{Wang:2022mrh}, by incorporating a monopole {\em Ansatz} as follows:
\begin{equation}\label{eqn::scalarpole}
\Delta_S^{\text{pole}}(Q^2)=\frac{\Delta_S(0)}{1+Q^2/m_S^2}\,,
\end{equation}
with $m_S$ being the scalar meson mass extracted from the corresponding BSE. Clearly, this correction only amends the $Q^2 > 0$ evolution of the scalar propagator. It will not alter the result for the $D-$term.
\par
For the sake of completeness, one could also consider the EFF, $F_{EM}(Q^2)$. Following an analogous procedure, $F_{EF}(Q^2)$ is obtained via the following expression:
\begin{equation}
\label{eq:EFFdef1}
K_\mu F_{EM}(Q^2)=N_c\text{tr}\int_q \gamma_\mu S(q+p)i^2F(q,p,-k) S(q+k)\,.
\end{equation}
Beyond its similarity to Eq.\,\eqref{eqn::emt}, a closer examination of Eq.\,\eqref{eq:EFFdef1} uncovers the typical form of the impulse approximation; thus, the tree-level quark-photon vertex becomes effectively dressed, featuring a vector meson pole in the time-like axis (see Refs.\,\cite{Roberts:2010rn,Gutierrez-Guerrero:2010waf}).

\section{results}\label{sec::results}
In the previous section we gathered the essential ingredients for computing the EMT valence-quark contribution, and the different components that it requires within the CI model. The parameters that define this model are the infrared mass scale $m_G$, which sets the effective strength of the interaction, the ultraviolet and infrared cutoffs ($\Lambda_{uv}$ and $\Lambda_{ir}$, respectively), and the current quark masses. The conventional way of setting these is such that the masses of the ground-state pseudoscalars are faithfully reproduced\,\cite{Gutierrez-Guerrero:2010waf,Roberts:2010rn,Roberts:2011cf}. The complete list of parameters is found in Table\,\ref{tab:parameters}. For the light quark sector, we have adopted widely used parameters found in early literature, \emph{e.g.} Refs.\,\cite{Yin:2019bxe,Gutierrez-Guerrero:2021rsx}. Whereas for heavy quark sector, we vary the parameters to obtain the correct meson masses and 
then confirm the robustness of our computations for the GFFs, obtaining the interaction radii and \textit{D}-term. 

\begin{table}[t]
\caption{\label{tab:parameters} CI model parameters employed herein, \ie, current quark mass $m_f$, gluon mass scale $m_G$, and ultraviolet regulator $\Lambda_{uv}$. In all cases, $\Lambda_{ir}=0.24\,\text{GeV}\sim \Lambda_{QCD}$. Mass units in GeV.}
\begin{tabular*}{0.45\textwidth}{c @{\extracolsep{\fill}} ccc}
\hline\hline
&$m_f$ &$m_G$ & $\Lambda_{uv}$\\
\hline
chiral limit &0 & 0.132 & 0.905\\
$u$&0.007 & 0.132 & 0.905\\
$s$& 0.17 & 0.132 & 0.905\\
\hline
&1.19 & 0.349512 & 2.0\\
&1.027& 0.461416 & 2.5\\
$c$&0.896& 0.568196 & 3.0\\
&0.791& 0.672299 & 3.5\\
&0.703& 0.772548 & 4.0\\
\hline
&3.86 &1.20569& 6.0\\
&3.53&1.41603& 7.0\\
$b$&3.24&1.62438& 8.0\\
&2.98&1.82809 & 9.0\\
&2.749&2.02682 & 10.0\\
\hline\hline
\end{tabular*}
\end{table}

\begin{table}[t]
\caption{\label{tab:quarkandmesons} Constituent quark and the $PS/S$ meson masses (in GeV).
For the $c$ and $b$ quarkonium sectors, the results are averaged for illustration purpose and the errors are determined through standard statistical methods. Note that chiral symmetry imposes $m_{PS}=0$ and $m_S=2M_f$ in the chiral limit\,\cite{Gutierrez-Guerrero:2010waf,Roberts:2010rn,Roberts:2011cf}.}
\begin{tabular*}{0.45\textwidth}{c @{\extracolsep{\fill}} ccc}
\hline\hline
&$M_f$ &$m_{PS}$ & $m_{S}$\\
\hline
chiral limit &0.3578 & 0.0& 0.7157\\
$u$&0.3677 & 0.1396 & 0.7416\\
$s$&  0.5331 & 0.7012 & 1.1364\\
$c_{avg.}$&$1.6241(40)$ & $2.9790(3)$ &  $3.4216(8)$\\
$b_{avg.}$&$4.8777(3)$ & $9.4012(8)$ &  $9.8604(6)$\\
\hline\hline
\end{tabular*}
\end{table}
\par
\begin{figure}
\includegraphics[width=8.6cm]{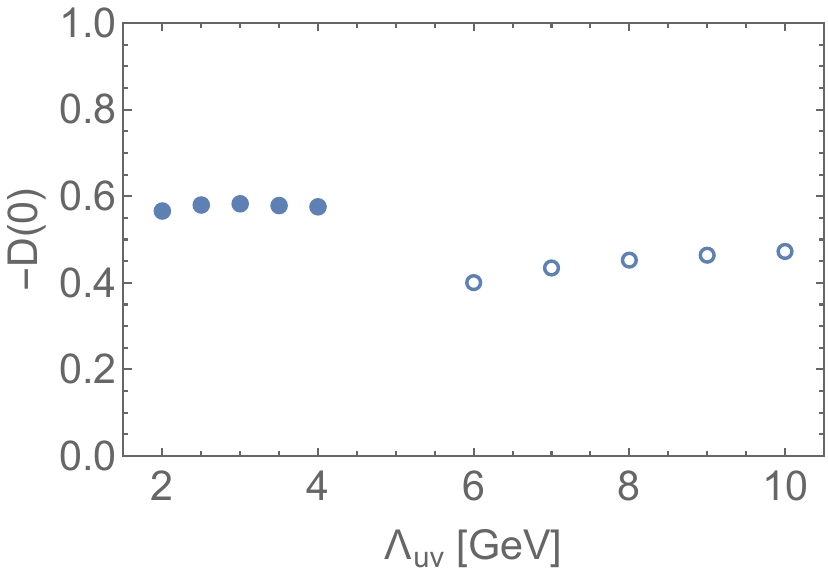}
\includegraphics[width=8.6cm]{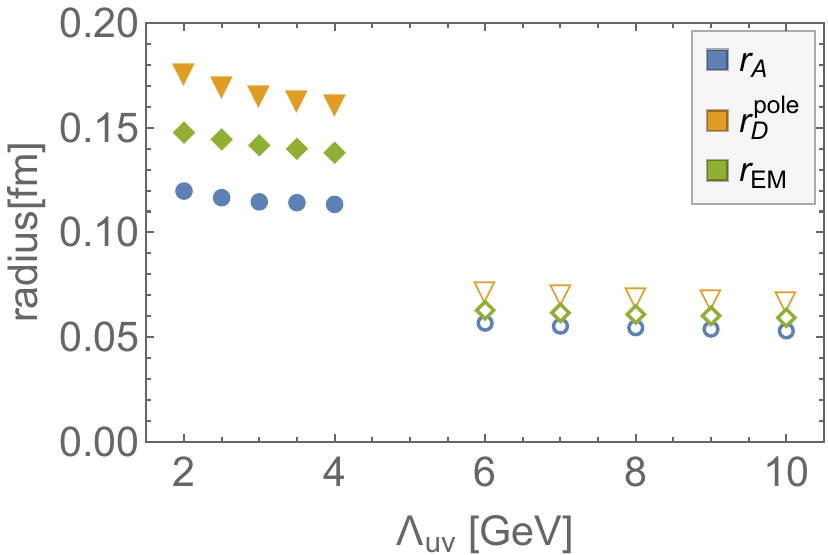}
\caption{\label{fig:uv_dependence}$\Lambda_{uv}$ dependence of the \textit{D}-term (upper panel), and the radii (lower panel) for $\eta_c$ (solid markers) and $\eta_b$ (open markers) mesons. Recall that for each value of $\Lambda_{uv}$, the interaction strength ($m_G$) and current quark masses are set to produce the observed value for $m_{PS}$.}
\end{figure}

Using the parameters listed in Table\,\ref{tab:parameters}, 
we first proceed to study the variation of the $D-$term with respect to the model inputs in the heavy quarkonia regime. In particular, the upper panel of \Fig{fig:uv_dependence} depicts the $\Lambda_{uv}$ dependence of the $D-$term. 
The robustness of the results with respect to the model parameters is further verified by computing the interaction radii associated with the GFFs and EFFs as follows:
\begin{equation}
\label{eq:radDef}
    \langle r_\#^2\rangle=-6\frac{\partial \text{ln}F_\#(Q^2)}{\partial Q^2}|_{Q^2=0}\,,
\end{equation}
such that the index $\#$ denotes the FF under study. The resulting variation of the mass ($r_A$), mechanical ($r_{D}$) and electromagnetic ($r_{EM}$) radii with respect to $\Lambda_{uv}$ is depicted in the lower panel of Fig.\,\ref{fig:uv_dependence}. Clearly, there is a  mild dependence on $\Lambda_{uv}$ which stabilizes with increasing $\Lambda_{uv}$ despite a considerably wide range of parameter selection shown in Table\,\ref{tab:parameters}. Therefore, the value of the $D-$term and the radii associated with the FFs are reasonably well-determined once the mass of the meson is accurately fixed. Although alternative definitions exist (e.g., Ref.~\cite{Polyakov:2018zvc}), we have chosen to designate the radii associated with the gravitational form factor $A(Q^2)$ and $D(Q^2)$ gravitational form factors as “mass” and “mechanical” radii, respectively, to keep the analogy with the charge radius and in order to facilitate comparisons between the different quantities. Such a choice has been employed, for instance, in the phenomenological analysis of Ref.\,\cite{Kumano:2017lhr}, and it is quite common within the DSE approach\,\cite{Xu:2023bwv,Xing:2022mvk,Raya:2024ejx}.

\begin{figure}
\includegraphics[width=8.6cm]{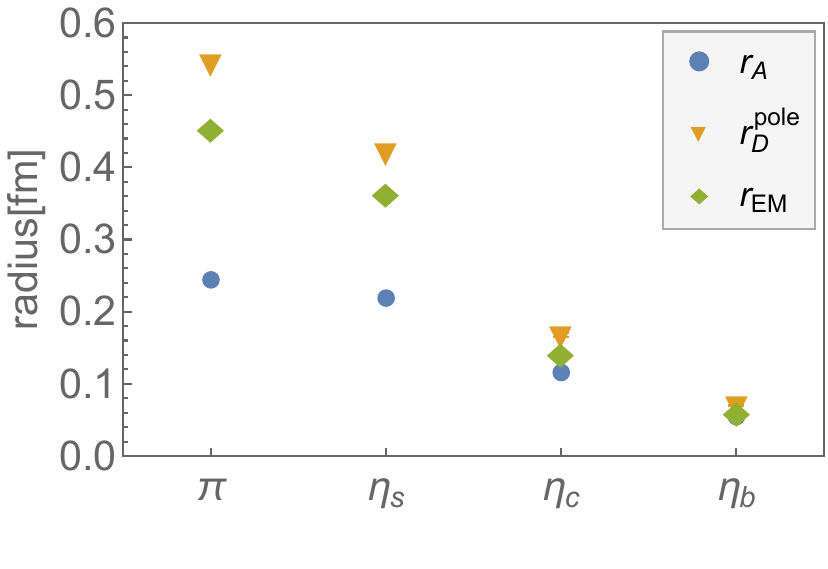}
\caption{\label{fig:Radii} Mass ($r_A$), mechanical ($r_D$) and  electromagnetic ($r_{EM}$) radii for different $PS$ mesons.  The error bands in the case of $\eta_c$ and $\eta_b$ mesons are practically negligible.}
\end{figure}

\begin{figure}
\includegraphics[width=8.6cm]{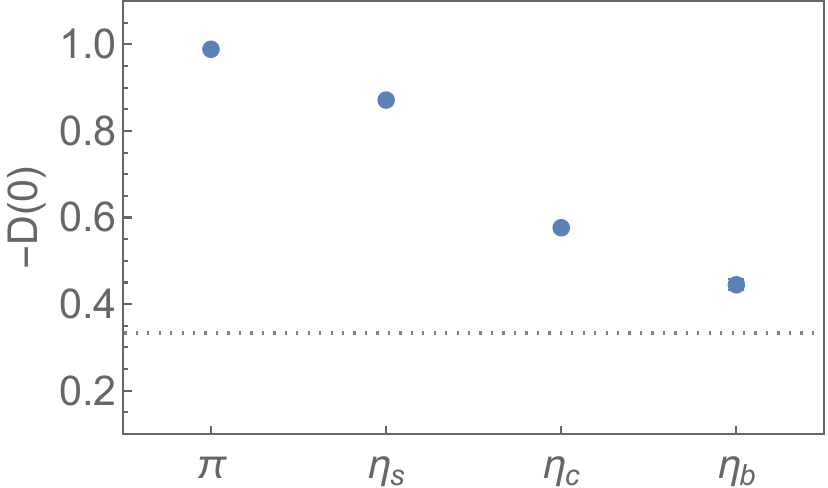}
\caption{\label{fig::Dterm}Analogous of \,\Fig{fig:Radii} for the pseudoscalar mesons $D-$term estimates.}
\end{figure}

After the model (in)dependence on the input has been tested, the mass, mechanical and  electromagnetic radii are depicted in\,\Fig{fig:Radii}. The error bars in the heavy quarkonia case are practically imperceptible due to the mild model dependence which also translates into small error bands in the associated dressed quark and the meson masses. The pion case has been considered  along with a hypothetical $s\bar{s}$ pseudoscalar, denoted by $\eta_s$, for comparative purposes and, therefore, no error estimates have been included. A similar comparison for the $D-$term estimates is shown in Fig.\,\ref{fig::Dterm}. Focusing on the interaction radii, there are some relevant points that require discussion. First of all, it is clear that these values follow the hierarchy:
\begin{equation}
\label{eq:radiiHie}
 r_A\,\text{(mass)}\, < r_{EM}\,\text{(charge)}\, < r^{\text{pole}}_D\,\text{(mechanical)}\,. 
\end{equation}
From various theoretical and phenomenological perspectives, this pattern has already been observed in the case of the pion, \emph{e.g.}\,\cite{Kumano:2017lhr,Hackett:2023nkr,Xing:2022mvk,Xu:2023izo,Raya:2021zrz,Broniowski:2024oyk,Xu:2023bwv,Wang:2024sqg}. However, the magnitude of the ratios $r_A/r_{EM}$ and $r_D/r_{EM}$ is still a matter of discussion\,\cite{Wang:2024sqg}. Notably, the hierarchy of interaction radii is maintained in the case of the heavier pseudoscalar mesons, but they become increasingly smaller and the $r_{A,D}/r_{EM}$ ratios approach unity as the mass of the bound state increases. This observation is intuitively expected\,\cite{Raya:2024ejx}, since, as the mass of the ground-state pseudoscalar grows, the bound state tends to feature point-like characteristics and its internal dynamics approaches the non-relativistic limit\,\cite{Neubert:2005mu}. In fact, considering the following averaged ratio
\begin{equation}
    \label{eq:ratAve}
    \bar{r}_H=\frac{1}{3}\sum_{\#} \frac{r_{\#}^H}{r_{\#}^\pi}\;,
\end{equation}
it is estimated that the spatial extent of the $\eta_c$ meson is approximately $36\,\%$ of that of the pion, whereas the corresponding for $\eta_b$ is around $16\,\%$. Being composed of relatively light quarks, the estimated size of the $\eta_s$ is only reduced about $20\,\%$ with respect of the pion. In the Table~\ref{tab:ratios}, we collect relevant values concerning the interaction radii: electromagnetic radii calculated for the pion, $\eta_c$ and $\eta_b$; the $r_A/r_{EM}$ and $r_D^{\text{pole}}/r_{EM}$ ratios, and the averages $\bar{r}_H$. Given the nature of the CI model, it is not surprising to find smaller interaction radii than those inferred from data or predicted by more sophisticated methods\,\cite{Gutierrez-Guerrero:2010waf,Roberts:2010rn,Roberts:2011cf}. However, since the $A(Q^2)$ GFF is entirely determined by the contribution of $\mathcal{M}_{1\mu\nu}$ component of the $HH$ scattering amplitude, corresponding to the impulse approximation, and this, in turn, does not possess the relevant resonance poles, the resulting ratio $r_A/r_{EM}$ is relatively small (of the order of $1/2$) and consistent with the large $N_c$ limit computation~\cite{Polyakov:1998ze}.

\begin{table}[t]
\caption{\label{tab:ratios}Electromagnetic radii (in fm) and the $r_{A,D}/r_{EM}$ ratios for different pseudoscalar mesons. We have also included the average $\bar{r}$ defined in Eq.\,\eqref{eq:ratAve}; the error in this case comes from the usual definition of the standard error. }
\begin{tabular*}{0.45\textwidth}{c| @{\extracolsep{\fill}} cccc}
\hline\hline
& $\pi$ & $\eta_s$ & $\eta_c$ & $\eta_b$ \\
\hline
$r_{EM}$& 0.455 &  0.364 & 0.142 & 0.061 \\
$r_A/r_{EM}$& 0.537 & 0.601 & 0.813 & 0.896 \\
$r_D^{\text{pole}}/r_{EM}$& 1.197 & 1.157 & 1.178 & 1.147 \\
$\bar{r}$ & 1 & 0.82(4) & 0.37(6) & 0.16(3) \\
\hline\hline
\end{tabular*}
\end{table}

Returning to the $D(0)$ estimations presented in Fig.\,\ref{fig::Dterm}, we also see a clear trend: its absolute value decreases as the mass of the pseudoscalar meson increases. In particular, including the estimated model uncertainties for the heavy quarkonia, we obtain:
\begin{eqnarray}
\nonumber
    D^{\pi} = -0.989&,&D^{\eta_s}= -0.871\,,\\
    \,D^{\eta_c}=-0.576(3)&,&D^{\eta_b}=-0.445(13)\,.\,\label{eq:DtermsM}
\end{eqnarray}
The result for the pion $D^{\pi}$ lies within the typical reported range\,\cite{Xu:2023izo,Polyakov:2018zvc,Hudson:2017xug,Broniowski:2024oyk}, and it is also consistent with the chiral limit constraint, Eq.\,\eqref{eq:sptc}. The approach employed here allows us to reproduce this limit exactly, while also predicting the corresponding chiral limit value for the scalar meson case:
\begin{equation}
    D^{PS}_{c.l.}=-1\,,\,D^{S}_{c.l.}=-\frac{7}{3}\,.
\end{equation}
As discussed previously, corrections to $D^{PS}_{c.l.}$ can be evaluated systematically using $\chi$PT, producing\,\cite{Polyakov:2018zvc,Donoghue:1991qv,Hudson:2017xug}: $D^{\pi}=-0.97(1)\,,\,D^{K}=0.77(15)$. These values turn out to be fully compatible with recent estimates based on DSEs\,\cite{Xu:2023izo}. Eq.\,\eqref{eq:DtermsM} and Fig.\,\ref{fig::Dterm} again depict a clear tendency to move away from $D^{PS}_{c.l.}=-1$ reducing its absolute value quadratically as $m_{PS}^2$ increases. This pattern stands in stark contrast to recent findings from the basis light-front quantization (BLFQ) approach, which suggests  $D^{PS,S}\sim -5$ in the charmonia sector,\,\cite{Xu:2024hfx}. Despite the valuable insights offered by the BLFQ computation, most likely, the current level of sophistication in calculating GFFs using such approach still requires further refinement.

In this context, it is noteworthy to mention the recent observations from Ref.\,\cite{Maynard:2024wyi}, where the GFFs are examined within a $\Phi^4$ theory in the small coupling regime. Therein, it is found that the one-loop corrections to the $D_{free}^{\Phi} = -1$ case reduce this value to $D_{1-loop}^{\Phi} = -1/3$. This result has been argued to be independent of the strength of the perturbative correction and could potentially be generalized to other theories. Acknowledging that the bound states tend to display more point-like characteristics with increasing mass (as revealed for instance, by the interaction radii discussed before), interpreting the increasingly heavier states as point-like particles become progressively more appropriate. Thus, the value $D_{1-loop}^{\Phi}$ could be considered as a limit for the infinitely massive case. This would mean that $D^{PS}$ would be bounded within the range:
\begin{equation}
\label{eq:Dbounds}
    D^{PS}\in (-1,-1/3)\,;
\end{equation}
where, clearly, $-1$ corresponds to the chiral limit and $-1/3$ to the infinitely massive case. The results presented here are plainly compatible with Eq.\,\eqref{eq:Dbounds} and, in addition, suggest the $D^S \in (-7/3,-1/3)$ bounds for the scalar mesons. Furthermore, the so-called weak energy condition resulting form the assumption of positive energy density,\,\cite{Xu:2024hfx}, entails:
\begin{equation}
    D^H \leq -1 + \frac{2}{3}m_H^2 r_A^2 \;.
\end{equation}
Evidently, the result $D^{PS}_{c.l.}=-1$ is properly recovered in the chiral limit and, given the values obtained for the quarkonia masses and radii (see Table\,\ref{tab:ratios} and \ref{tab:quarkandmesons}), it is easy to show that this inequality is satisfied.

\par
\begin{figure}
\includegraphics[width=8.6cm]{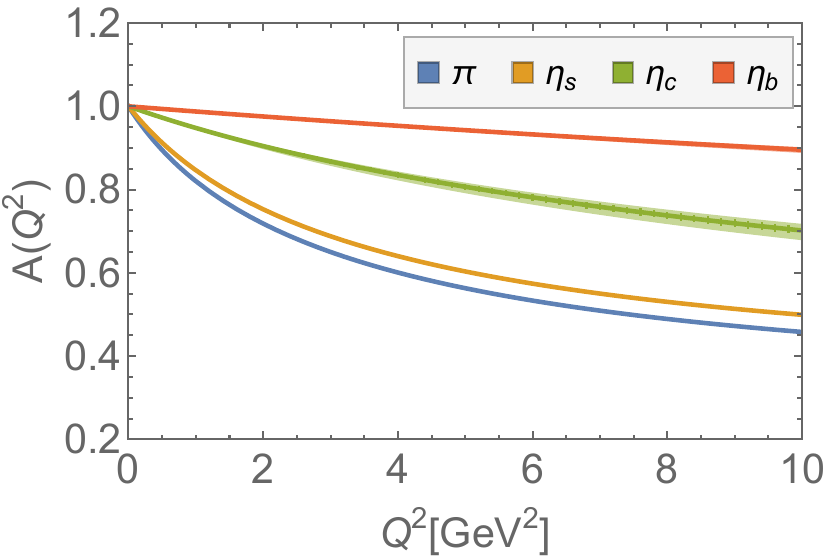}
\includegraphics[width=8.6cm]{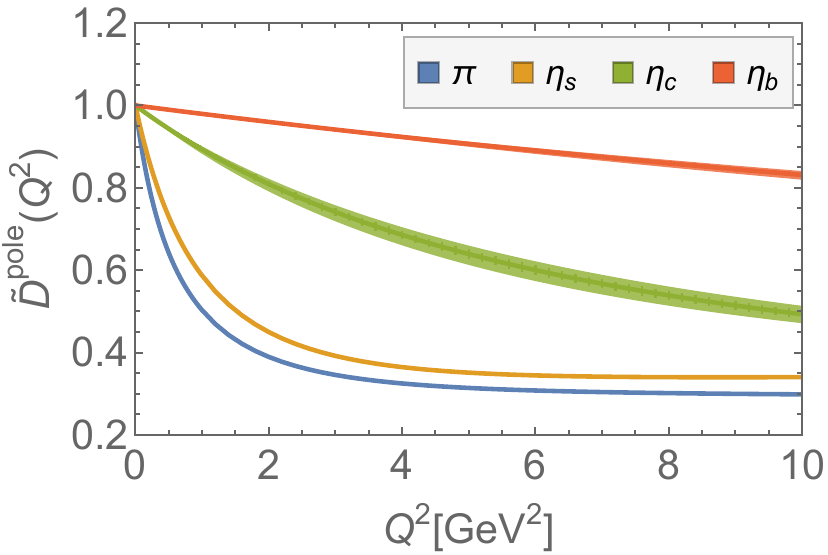}
\caption{\label{fig:GFF}GFFs of the $PS$ mesons. The $D(Q^2)$ FF is normalized as $\Tilde{D}(Q^2)=D(Q^2)/D(0)$ for a direct visual comparison. The error bands are calculated from the variation of the parameters listed in Table\,\ref{tab:parameters}.}
\end{figure}

The $Q^2$ dependence of $A(Q^2)$ and $D^{\text{pole}}(Q^2)$  GFFs of the $PS$ light and heavy quarkonia are shown in Fig.\,\ref{fig:GFF} Only a slight dependence on the model parameters is observed; this essentially corresponds to the variation of $\Lambda_{uv}$, since the rest of the parameters are set accordingly. The FFs are monotonically decreasing functions of $Q^2$, albeit with a less pronounced curvature as would be expected within more realistic models\,\cite{Xu:2023izo}. As we have seen, the low-$Q^2$ region allow us to determine the corresponding interaction radii (see Eq.\,\eqref{eq:radDef}), while $D(Q^2=0)$ defines the $D-$term. The hierarchical order of the interaction radii exposed in Eq.\,\eqref{eq:radiiHie} is naturally reflected in the hardness of the FFs: the $A(Q^2)$ GFF turns out to be above the EFF, which in turn is above $D(Q^2)$. Obviously, when categorizing by meson type, the heavier ones produce FFs with a less pronounced curvature, becoming more and more aligned with a point-like particle picture. 

We recall that $A(Q^2)$ entirely emerges from the $\mathcal{M}_{1\mu\nu}$ contribution to the $HH$ scattering amplitude, defining the impulse approximation piece. Clearly, this is enough to guarantee the mass normalization condition, $A(0)=1$. On the other hand, the $D(Q^2)$ GFFs do require the systematic addition of the $\mathcal{M}_{2\mu\nu}$ component, which features the scalar meson pole. This turns out necessary to ensure the soft-pion theorem constraint is faithfully reproduced, hence yielding a proper normalization of $D(Q^2)$ (the $D-$term).  

\section{Summary and perspectives}\label{sec::summary}

We calculate the GFFs of pion and $PS$ quarkonia $\eta_s$, $\eta_c$ and $\eta_b$, within the coupled DSE-BSE formalism. The cornerstone of this calculation is the systematic construction and dressing of the corresponding $HH$ amplitude, in contrast with the typical approach of constructing the QTV.  This novel procedure was suggested in Ref.\,\cite{Xing:2022jtt}. It requires the knowledge of the quark propagator, the meson BSA as well as the incorporation of an effective tree-level like QTV. These defining elements are constructed within the CI model which is widely used in the literature since it was proposed initially in Refs.\,\cite{Gutierrez-Guerrero:2010waf,Roberts:2010rn,Roberts:2011cf}. The CI framework is not only computationally amicable but is also highly illustrative, since the origin and identification of the outcomes is easily discernible. Remarkably, the approach employed herein makes it possible to produce the GFFs that satisfy the known constraints. We also calculate the relevant interaction radii (mass, electromagnetic and mechanical) and the \textit{D}-term, which is obtained by setting $Q^2 = 0$ and provides valuable insights into the underlying strong interaction dynamics.
\par
Given the apparent freedom to fix the model parameters, we firstly study the robustness of our results as a function of their variation. For this purpose, we let the ultraviolet mass scale $\Lambda_{uv}$ take values within a reasonable range of hadronic mass scales, and fix the remaining free parameters (coupling constant $m_G$ and current quark mass) to produce the experimental ground-state $PS/S$ meson mass under consideration. Both the interaction radii and the $D-$term turn out to be only weakly sensitive to this variation. So their values are practically dictated by the mass of the meson. 
\par
In the case of interaction radii, we note the following hierarchy for $PS$ mesons: $r_A < r_{EM} < r^{\text{pole}}_D$. This pattern is consistent with several explorations in the light sector, but the specific values of these radii are currently subject to scrutiny (see\,\cite{Wang:2024sqg} for a recent discussion). We verify that the $D-$term moves away from the expectations in the chiral limit, $D^{PS}_{c.l.}=-1$, reducing its absolute value as the mass of the $PS$ meson increases; specifically, we find that $D^{\eta_s} \approx -0.87$, $D^{\eta_c}\approx - 0.58$ and $D^{\eta_b}\approx-0.45$. This trend is reinforced by $\chi$PT expectations for the NG bosons\,\cite{Hudson:2017xug,Donoghue:1991qv,Polyakov:2018zvc}. Together with this observation and the fact that our formalism exactly satisfies the relevant symmetries (among others, mass normalization, charge conservation, and soft-pion theorem), we believe it to be a reliable indicative that sensible values have been obtained for the $D-$term of the quarkonia systems studied in this article. 
\par
Furthermore, recent analysis of a  $\Phi^4$ theory at one loop level~\cite{Maynard:2024wyi}, has revealed that the $D-$term of a free particle is significantly reduced after a one-loop correction; in particular, it is found that $D_{1-loop}^\Phi=-1/3$, regardless of the coupling constant. It may potentially be a general result in scalar theories. In QCD, this value can be interpreted as a limit for an infinitely massive $PS/S$ system, corresponding to the case in which the bound-state would manifest the characteristics of a point particle. Therefore, both the latter bound and the soft-pion theorem would impose upper and lower limits on the values that the $D-$term of $PS$ mesons can take, \emph{i.e.} $D^{PS}\in (-1,-1/3)$. Our symmetry-preserving calculations are consistent with this expectation.
\par
As for the scalar mesons, our model and the computational scheme yield $D^S_{c.l.}=-7/3\approx -2.33$ in the chiral limit. For the arguments detailed above, we anticipate the range of possible values: $D^S \in (-7/3,-1/3)$. As a promising comparison, the Nambu-Jona-Lasinio model predicts $D^{\sigma}=-2.27$,\,\cite{Freese:2019bhb}; this result perfectly agrees with our prediction. Note that despite the straightforward extension to scalar mesons due to the characteristic simplicity of the CI model, we have not charted out the $Q^2$ evolution of the GFFs of scalar mesons. 
It is due to the fact that a refined determination of the GFFs for the scalar mesons would potentially reveal a richer internal structure beyond a quark-antiquark picture; for instance, unveiling a greater coupling with the surrounding meson cloud, and the formation of tetraquark-like structures. Thus any attempt to compute these form factors is not merely an algebraic  exercise but also provides a window into an intriguing complexity and richer structure of the strong interactions at the level of quarks and the bound states. Finally, remembering that beyond $D^{PS}_{c.l.}=-1$ there is no known prescription for the hadronic $D-$term, the importance of its determination for any system is self-evident. In addition to other theoretical endeavors, we also hope that lQCD and modern experiments will be able to provide confirmatory/reliable answers.

\section{Acknowledgments}
This work is supported by the National Natural Science Foundation of China (grant no. 12135007),
the Spanish MICINN grant PID2022-140440NB-C22, and the
regional Andalusian project P18-FR-5057. A. B. acknowledges
Coordinaci\'on de la Investigaci\'on Cient\'ifica, Universidad
Michoacana de San Nicol\'as de Hidalgo grant 4.10. A. B. also wishes to thank Beatriz-Galindo support during his current stay at the University of Huelva, Huelva, Spain.

\bibliography{apstemplateNotes}
\end{document}